\documentclass{article}

\usepackage{amsmath, amsthm, amscd, amsfonts, amssymb, graphicx, color}
\usepackage[bookmarksnumbered, colorlinks, plainpages]{hyperref}
\usepackage{pstricks}
\usepackage{pspicture,pst-all,multido}
\usepackage[utf8]{inputenc}

\theoremstyle{definition}

\theoremstyle{remark}

\numberwithin{equation}{section}

\newcommand{\e}{\operatorname{e}}

\renewcommand{\i}{\operatorname{i}}

\renewcommand{\d}{\operatorname{d}}

\newcommand{\C}{\mathbb{C}}
\newcommand{\R}{\mathbb{R}}

\newcounter{envcount}%

{\vspace{\bigskipamount}\refstepcounter{envcount}\textbf{(\theenvcount)\enspace Remarks.}}%
  {\vspace{\bigskipamount}}

{\vspace{\bigskipamount}\refstepcounter{envcount}\textbf{(\theenvcount)\enspace Notation.}}%
  {\vspace{\bigskipamount}}

{\vspace{\bigskipamount}\refstepcounter{envcount}\textbf{(\theenvcount)\enspace Limit hyperplanes.}}%
  {\vspace{\bigskipamount}}

{\vspace{\bigskipamount}\refstepcounter{envcount}\textbf{(\theenvcount)\enspace Remark on}}%
  {\vspace{\bigskipamount}}

{\vspace{\bigskipamount}\refstepcounter{envcount}\textbf{(\theenvcount)\enspace Remark.}}%
  {\vspace{\bigskipamount}}

{\vspace{\bigskipamount}\refstepcounter{envcount}\textbf{(\theenvcount)\enspace Properties of achronal sets.}}%
  {\vspace{\bigskipamount}}

{\vspace{\bigskipamount}\refstepcounter{envcount}\textbf{(\theenvcount)\enspace Properties of causal bases.}}%
  {\vspace{\bigskipamount}}

{\vspace{\bigskipamount}\refstepcounter{envcount}\textbf{(\theenvcount)\enspace Definition.}}%
  {\vspace{\bigskipamount}}

{\vspace{\bigskipamount}\refstepcounter{envcount}\textbf{(\theenvcount)\enspace POL for Weyl particles.}}%
  {\vspace{\bigskipamount}}

{\vspace{\bigskipamount}\refstepcounter{envcount}\textbf{(\theenvcount)\enspace Summary.}}%
  {\vspace{\bigskipamount}}

{\vspace{\bigskipamount}\refstepcounter{envcount}\textbf{(\theenvcount)\enspace Time Reversal.}}%
  {\vspace{\bigskipamount}}
  
\newenvironment{L-CS}%
{\vspace{\bigskipamount}\refstepcounter{envcount}\textbf{(\theenvcount)\enspace Large-\,$t_{\overline{e}}$\,-states.}}%
  {\vspace{\bigskipamount}}

{\vspace{\bigskipamount}\refstepcounter{envcount}\textbf{(\theenvcount)\enspace Asymptotic causality.}}%
  {\vspace{\bigskipamount}}

{\vspace{\bigskipamount}\refstepcounter{envcount}\textbf{(\theenvcount)\enspace Example.}}%
  {\vspace{\bigskipamount}}

{\vspace{\bigskipamount}\refstepcounter{envcount}\textbf{(\theenvcount)\enspace Example}}%
  {\vspace{\bigskipamount}}

{\vspace{\bigskipamount}\refstepcounter{envcount}\textbf{(\theenvcount)\enspace Construction}}%
  {\vspace{\bigskipamount}}

{\vspace{\bigskipamount}\refstepcounter{envcount}\textbf{(\theenvcount)\enspace Discussion. }}%
  {\vspace{\bigskipamount}}  

\newenvironment{The}%
{\vspace{\bigskipamount}\refstepcounter{envcount}\textbf{(\theenvcount)\enspace Theorem.}\itshape}%
  {\vspace{\bigskipamount}\upshape}

{\vspace{\bigskipamount}\refstepcounter{envcount}\textbf{(\theenvcount)\enspace Theorem}\itshape}%
  {\vspace{\bigskipamount}\upshape}
  
\newenvironment{Pro}%
{\vspace{\bigskipamount}\refstepcounter{envcount}\textbf{(\theenvcount)\enspace Proposition.}\itshape}%
  {\vspace{\bigskipamount}\upshape}

\newenvironment{Cor}%
{\vspace{\bigskipamount}\refstepcounter{envcount}\textbf{(\theenvcount)\enspace Corollary.}\itshape}%
  {\vspace{\bigskipamount}\upshape}

{\vspace{\bigskipamount}\refstepcounter{envcount}\textbf{(\theenvcount)\enspace Frame-dependence of NWL.}\itshape}%
  {\vspace{\bigskipamount}\upshape}

{\vspace{\bigskipamount}\refstepcounter{envcount}\textbf{(\theenvcount)\enspace Construction of a late-change state.}\itshape}%
  {\vspace{\bigskipamount}\upshape}

\newenvironment{Lem}%
{\vspace{\bigskipamount}\refstepcounter{envcount}\textbf{(\theenvcount)\enspace Lemma.}\itshape}%
  {\vspace{\bigskipamount}\upshape}

{\vspace{\bigskipamount}\refstepcounter{envcount}\textbf{(\theenvcount)\enspace Example.}\itshape}%
  {\vspace{\bigskipamount}\upshape}

{\vspace{\bigskipamount}\refstepcounter{envcount}\textbf{(\theenvcount)\enspace Example}\itshape}%
  {\vspace{\bigskipamount}\upshape}

{\vspace{\bigskipamount}\refstepcounter{envcount}\textbf{(\theenvcount)\enspace Lorentz contraction.}\itshape}%
  {\vspace{\bigskipamount}\upshape}

  {\vspace{\bigskipamount}}

{\vspace{\bigskipamount}\refstepcounter{envcount}\textbf{(\theenvcount)\enspace Interpretation.}\itshape}%
  {\vspace{\bigskipamount}\upshape}

{\vspace{\bigskipamount}\refstepcounter{envcount}\textbf{(\theenvcount)\enspace Canonical cross section.}}%
  {\vspace{\bigskipamount}}

\newenvironment{Po}%
{\vspace{\bigskipamount}\refstepcounter{envcount}\textbf{(\theenvcount)\enspace Proof of}}%
  {\vspace{\bigskipamount}}

\theoremstyle{definition}
\swapnumbers

\theoremstyle{remark}

\begin{document}
\setcounter{page}{1}
\pagenumbering{arabic}

\vspace{2mm}
\begin{center}
{\Large Localization of the massive scalar boson on achronal hyperplanes, derivation of Lorentz contraction}\\  

\vspace{1cm}
Domenico P.L. Castrigiano\\
Technische Universit\"at M\"unchen, Fakult\"at f\"ur Mathematik, M\"unchen, Germany\\ 

\smallskip

{\itshape E-mail address}: {\tt
castrig\,\textrm{@}\,ma.tum.de}\\

\end{center}

%\author[D.P.L. Castrigiano]{Domenico P.L. Castrigiano$^*$}
%\address{$^{*}$Technische Universit\"at M\"unchen, Fakult\"at f\"ur Mathematik, M\"unchen, Germany}
%\email{\textcolor[rgb]{0.00,0.00,0.84}{castrig@ma.tum.de}}
\abstract

It is shown that the causal localizations of the massive scalar boson on spacelike hyperplanes extend uniquely to all achronal hyperplanes. The extension occurs by means of the high boost limit in a  covariant manner. Towards a localization in maximal achronal   surfaces a simple but  emblematic case shows that normalization, demanded by causality, is preserved. Moreover the existence of the high boost limit, as a consequence of causality, implies the phenomenon of Lorentz contraction discussed in detail. In conclusion, these considerations constitute a clear plea for the concept of achronal localization.

\section{Introduction}
We investigate an emblematic consequence of causality regarding  localizable quantum mechanical systems.  It yields an irrefutable argument why to extend   spacelike localization to achronal localization. Invoking further  physical grounds it is argued in   \cite{C24} that a Poincar\'e covariant \textbf{achronal localization}
constitutes  the frame which complies most completely with  the principle of causality for quantum mechanical systems.
Actually it is equivalent to a covariant representation of the causal logic \cite[(21)]{C24}.
\\
\hspace*{6mm}
Commonly localizability is described by a Poincar\'e covariant positive operator valued map $T$ being a normalized measure on every spacelike hyperplane of Minkowski space. The expectation value of the localization operator $T(\Delta)$ indicates the localization probability of the quantum system in the flat spacelike region $\Delta$. $T$ is \textbf{normalized} in that it asigns the unit operator to every spacelike hyperplane.
\\
\hspace*{6mm}
\textbf{Causality} imposes on the  localization $T$ the condition that  the probability of localization in a region of influence $\Delta'$ is not less than that in the region  of actual localization  $\Delta$ (\ref{CCFR}),  \cite[sec.\,11]{C23}. 
\\
\hspace*{6mm}
This causality condition implies a remarkable property of $T$. In the limit of infinite rapidity a  spacelike hyperplane,  boosted along a direction parallel to it, equals  a tangent  space of a light cone. This  tangent hyperplane  is no longer spacelike but still achronal, which means that the Minkowski distance of any two of its points is  not timelike. The limit is called \textbf{high boost limit} if it occurs pointwisely   such that every point of spacetime runs along a lightlike straight line. In this case the probabilities of localization converge. This is the property of  a causal localization $T$ alluded to above.
\\
\hspace*{6mm}
In the cases of the Dirac fermions electron and positron and of the four Weyl fermions it is shown in \cite{C17} that by the high boost limit the localization $T$ extends in a covariant manner to all achronal hyperplanes. The present investigations show that this property holds true also for the causal localizations of the massive scalar boson \cite{C23}.
\\
\hspace*{6mm}
So roughly speaking  by continuity every causal localization automatically comprises the regions of all achronal hyperplanes  preserving  Poincar\'e covariance.
\\
\hspace*{6mm}
The obvious question is how to extend the localization $T$ to maximal  achronal surfaces composed by flat  pieces. Achronal localization demands \textbf{normalization} in order to comply with causality \cite[(16) Theorem]{C24}. In (\ref{AET}) normalization is shown in a simple but emblematic case. It is provided just by the surplus of  probability of localization in the region of influence with respect to the initial region due to the requirement of causality.
\\
\hspace*{6mm}
The  aim, of course, is the extension of the localization to all maximal  achronal surfaces. Actually, in their 
thorough study \cite{DM24} on the subject  De C. Rosa, V. Moretti  succeed in extending $T$  to all  differentiable \textbf{Cauchy surfaces} maintaining normalization. Moreover, their methods promise a successful approach to the final goal.
\\
\hspace*{6mm}
The more striking  physical consequence of the existence of the high boost limit due to causality is the \textbf{Lorentz contraction} of the massive scalar boson. If boosted with sufficiently high rapidity the boson, in the boosted state $\phi'$,  is almost strictly localized in a whatever narrow strip  $S$ perpendicular to direction of the boost, i.e., the probability of localization satisfies
$$P_{\phi'}(S) \lessapprox 1$$
The questions regarding the ascertainment of  the Lorentz contraction are discussed in detail.

\section{Notations and notions} 
Vectors in $\R^4$ are denoted  by  $\mathfrak{x}=(x_0,x)$ with $x:=(x_1,x_2,x_3)\in\R^3$. Let $\varpi:\R^4\to\R^3$ denote the projection $\varpi(\mathfrak{x}):=x$.  Representing Minkowski spacetime by $\R^4$ the Minkowski product of $\mathfrak{a},\mathfrak{a}'\in\R^4$ is given by $\mathfrak{a}\cdot \mathfrak{a}':=a_0a_0'-aa'$, where for vectors $a,a'$ in $\R^3$ the scalar product $a_1a'_1+a_2a'_2+a_3a'_3$ is denoted by $aa'$. Often we use the notation $\mathfrak{a}^{\cdot 2}:=\mathfrak{a}\cdot\mathfrak{a}$. 
\\
\hspace*{6mm}
$\tilde{\mathcal{P}}=ISL(2,\C)$  is the universal covering group 
of  the Poincar\'e group.     
$\tilde{\mathcal{P}}$ acts on $\R^4$ as
\begin{equation}\label{PTUCH} 
g\cdot \mathfrak{x}:=\mathfrak{a}+\varLambda(A) \mathfrak{x}\quad  \text{ for } g=(\mathfrak{a},A)\in\tilde{\mathcal{P}}, \, \mathfrak{x}\in \R^4
\end{equation}
where $\varLambda:SL(2,\C)\to O(1,3)_0$ is  the universal covering homomorphism onto the proper orthochronous Lorentz group. For short one writes  $A\equiv (0,A)$, 
$\mathfrak{a} \equiv (\mathfrak{a},I_2)$, and $A\cdot \mathfrak{x}=\varLambda(A)\mathfrak{x}$. 
For $M\subset\R^4$ and $g\in \tilde{\mathcal{P}}$ define $g\cdot M:=\{g\cdot \mathfrak{x}: \mathfrak{x}\in M\}$.\\
\hspace*{6mm}
The group operation on  $\tilde{\mathcal{P}}$ reads 
$(\mathfrak{a},A)(\mathfrak{a}',A')
=(\mathfrak{a}+A\cdot\mathfrak{a}',AA')$ with identity element $(0,I_2)$ and inverse  $(\mathfrak{a},A)^{-1}=(-A^{-1}\cdot \mathfrak{a},A^{-1})$.
\\
 \hspace*{6mm}
A set $A\subset \R^4$ is said to be achronal if $|x_0-y_0| \le |x-y|$ for $\mathfrak{x},\mathfrak{y} \in A$. By definition $A$ is maximal achronal if $A$ in not properly contained in an achronal set. An achronal set  is maximal achronal if and only if it meets every timelike straight line.
 \\
\hspace*{6mm}
 The scalar boson with mass $m>0$ is described by the  mass shell representation 
 $W$  of   $\tilde{\mathcal{P}}$ on   $L^2(\mathcal{O})$. Here  $\mathcal{O}:=\{\mathfrak{p}\in\R^4:p_0=\epsilon(p)\}$,  $\epsilon(p):=\sqrt{m^2+p^2}$, is  the mass shell equipped with the Lorentz invariant measure $\d o(\mathfrak{p})\equiv \d^3 p/\epsilon(p)$. Explicitly\footnote{Often one uses the antiunitarily equivalent $ \e^{-\i \mathfrak{a}\cdot \,\mathfrak{p}}$.}

\begin{itemize}
\item $\big(W(\mathfrak{a},A)\phi\big)(p)= \e^{\i \mathfrak{a}\cdot \,\mathfrak{p}} \,\phi(A^{-1}\cdot \mathfrak{p})$
\end{itemize}

\section{Conserved covariant current} \label{CCC}

 As known the localizability of the massive scalar boson is described  by  a Euclidean covariant  positive operator valued measure  $T$ on the Borel sets of $\R^3$ such that the probability of localization in the region $\Delta$ of the boson in the state $\phi$ is supposed to be the expectation value  $\langle \phi,T(\Delta)\phi\rangle$ of the localization operator $T(\Delta)$. By  \cite[(6.1), (11)\,Theorem, (8.3)]{C23} one has
\begin{itemize} 
\item $\langle \phi,T(\Delta)\phi\rangle=\int_\Delta J_0(\phi,x) \d^3 x$
\end{itemize} 
 where the density of the probability of localization $J_0$ is given by
\begin{itemize}
\item $J_0(\phi,x):=(2\pi)^{-3}\int\int \, \textsc{k}(k,p) 
\e^{\i (p-k)x}\overline{\phi(\mathfrak{k})}\phi(\mathfrak{p})   \,\d o(\mathfrak{k}) \d o(\mathfrak{p})$
\end{itemize}
for  $\phi\in C_c$, i.e., continuous with compact support. 
Here  $\textsc{k}$ is a measurable rotational invariant positive definite  separable kernel  $\textsc{k}$ on $\R^3\setminus\{0\}$ with $\textsc{k}(p,p)=\epsilon(p)$. 
\\
\hspace*{6mm}
Due to the Euclidean covariance  $T(g\cdot\Delta)=W(g)T(\Delta) W(g)^{-1}$  for $g=(b,B)\in ISU(2)$, the localization  $T$  extends uniquely to all (Lebesgue) measurable spacelike flat regions $\Delta\subset\R^4$ in a Poincar\'e covariant manner 
and vanishes  just at the Lebesgue null sets  \cite[(9)]{C17}.
\\
\hspace*{6mm}
The intention  is to extend eventually  the localization  $T$ to all achronal spacetime regions of Minkowski space in order to achieve  a localization of the boson which complies in full with causality (cf.\,\cite{C24}). A promising way is in recognizing  $J_0$ to be the zero component of a conserved covariant four-vector current $\mathfrak{J}:=(J_0,J)$. Petzold and collaborators \cite{GGP67} show that this is the case if and only if
 
\begin{equation}\label{FVCCD}
\mathfrak{J}(\phi,\mathfrak{x})=(2\pi)^{-3}\int\int \, \frac{\mathfrak{k}+\mathfrak{p}}{2} g(\mathfrak{k}\cdot\mathfrak{p})
\e^{\i(\mathfrak{k}-\mathfrak{p})\cdot \mathfrak{x}}\overline{\phi(\mathfrak{k})}\phi(\mathfrak{p})   \, \d o(\mathfrak{k}) \d o(\mathfrak{p})
\end{equation}
where   $g: [m^2,\infty[ \to \R$ is  continuous with $g(m^2)=1$ such that  $(k,p)\mapsto (\epsilon(k)+\epsilon(p)) g(\mathfrak{k}\cdot\mathfrak{p})$ is a positive definite kernel on $\R^3$ (see also \cite[(55)\,Corollary]{C23}). For a thorough analysis of the solutions $g$ see \cite{C23}. We mention  $|g(t)|\le g_{3/2}(t)$, where $g_r(t):=(2m^2)^r(m^2+t^2)^{-r}$ for $r\ge 3/2$ denotes the basic series of solutions revealed by \cite{GGP67} and \cite{HW71}. Henceforth we will deal with conserved covariant  currents $\mathfrak{J}$ with positive definite kernel  (\ref{FVCCD}). 
\\
\hspace*{6mm}
$\mathfrak{J}(\phi, \cdot)$ is bounded and smooth since  $\phi\in C_c$.  Moreover $\mathfrak{J}$ satisfies  for all $\mathfrak{x}\in\R^4$,  $g\in\tilde{\mathcal{P}}$ (see  \cite[(52)\,Theorem]{C23})
\begin{itemize}
\item[(a)] $\operatorname{div} \mathfrak{J}(\phi,\mathfrak{x})=0$, i.e., the continuity equation
\item[(b)] $J_0(\phi,\mathfrak{x})\ge |J(\phi,\mathfrak{x})|$, i.e., $ \mathfrak{J}$ is zero or causal future-directed
\item[(c)] $\mathfrak{J}\big(W(g)\phi,\mathfrak{x}\big)=A\cdot\mathfrak{J}\big(\phi,g^{-1}\cdot\mathfrak{x}\big)$ Poincar\'e covariance
\end{itemize}

\section{Integration on achronal sets}

Let $\Lambda\subset\R^4$ be a  maximal achronal set determined by the differentiable  $1$-Lipschitz map $\tau:\R^3\to \R$ with $\Lambda =\{ \big(\tau(x),x\big): x\ \in \R^3 \}$ \cite{CJ79}, \cite[(1)\,(g)]{C24}. Let $\Delta$ be a Borel subset of $\Lambda$. The common formula for the flux through $\Delta$ by the vector field  $\mathfrak{J}$ reads
\begin{equation}\label{DETAS}
\pi_{\phi,\Lambda}(\Delta):=\int_{\varpi(\Delta)}\big(J_0(\phi,\tau(x),x)-J(\phi,\tau(x),x)\operatorname{grad}\tau(x)\big)\d^3x
\end{equation}
By (b) in sec.\,\ref{CCC} and since $|\operatorname{grad}\tau(x)|\le1$, (\ref{DETAS})  defines  a $\sigma$-additive measure  $\pi_{\phi,\Lambda}$ on $\Lambda$. The idea is that $\pi_{\phi,\Lambda}$ furnishes the desired extension of $T$ to the  achronal Borel sets by equating
\begin{equation}\label{TAS}
\langle \phi,T(\Delta)\phi\rangle=\pi_{\phi,\Lambda}(\Delta)
\end{equation}
This idea  is suggested by three reasons. (i) It is  
the very principle of causality which  let one think  of  the probability of localization as a conserved quantity reigned by an associated  density current. (ii) Equation (\ref{TAS})  holds for all spacelike hyperplanes $\Lambda$.
Actually,  \cite{DM24}  extends $T$ via (\ref{TAS}) to all  Borel subsets of differentiable Cauchy surfaces. Recall that a Cauchy surface is a set which meets every inextendible timelike smooth curve exactly once. Due to a result of V. Moretti \cite[Appendix D]{C24} a Cauchy surface turns out just to be a maximal achronal set, which intersect every lightlike straight line.  (iii) Last not least there is the covariance (\ref{CPOL})(b).

\begin{Pro}\label{CPOL} Let $g=(\mathfrak{a},A)\in \tilde{\mathcal{P}}$. Then 
\\
\hspace*{6mm}
\emph{(a)}  the achronal set $g\cdot\Delta$ equals  $\{(\tau_g(y),y):y\in \varpi(g\cdot\ \Delta)\}$ for $\tau_g(y):=(g\cdot (\tau(x),x))_0$ with $x:=S^{-1}(y)$, where $S:\R^3\to \R^3$, $S(x):=\varpi \big(g\cdot (\tau(x),x) \big)$ is a bijection.
\\
\hspace*{6mm}
\emph{(b)} $\pi_{W(g)^{-1}\phi,\Lambda}(\Delta)=\pi_{\phi,g\cdot\Lambda}(g\cdot\Delta)$
\end{Pro}\\
{\it Proof.} (a) Obviously $S$ is surjective. Let $S(x)=S(x')$. Hence $\varpi \big(g\cdot (\tau(x)-\tau(x'),x-x')\big)=0$ with  $\big(g\cdot (\tau(x)-\tau(x'),x-x')\big)^{\cdot 2}=\big(\tau(x)-\tau(x'),x-x'\big)^{\cdot 2}\le 0$. Therefore also $(g\cdot (\tau(x)-\tau(x'),x-x'))_0=0$, whence  $(g\cdot (\tau(x)-\tau(x'),x-x'))=0$. This means  $(\tau(x)-\tau(x'),x-x')=0$. So $x=x'$. In conclusion $S$ is bijective.
\\
\hspace*{6mm}
Note $S(\varpi(\Delta))=\varpi(g\cdot\Delta)$. Therefore $\{(\tau_g(y),y):y\in \varpi(g\cdot\ \Delta)\}=\{\big((g\cdot (\tau(x),x))_0, S(x)\big):x\in\varpi(\Delta)\} =\{g\cdot (\tau(x),x):x\in\varpi(\Delta)\}=g\cdot\Delta$.
\\
\hspace*{6mm}
(b) By (c)  sec.\,\ref{CCC},  $\pi_{ W(g)^{-1}\phi,\Lambda}(\Delta)=\int_{\varpi(\Delta) }\mathfrak{J}\big(\phi,g\cdot (\tau(x),x)\big)
 \cdot \big(A\cdot (1,\operatorname{grad}\tau(x))\big)\d^3x = \int_{\varpi(\Delta) }\mathfrak{J}\big(\phi, \tau_g(S(x)),S(x)\big)
 \cdot \big(A\cdot (1,\operatorname{grad}\tau(x))\big)\d^3x=    \int_{S(\varpi(\Delta)) }\mathfrak{J}\big(\phi, \tau_g(y),y\big)
 \cdot \big(A\cdot (1,\operatorname{grad}\tau(S^{-1}(y))\big)\d S(\lambda)(y)$, where $\lambda$ is the Lebesgue measure. Recall  $S(\varpi(\Delta))=\varpi(g\cdot\Delta)$ and note $\d S(\lambda)/\d \lambda =|\det D\,S^{-1}|=|\det D\,S(S^{-1}(\cdot))|^{-1}$.
\\
\hspace*{6mm} 
It remains to verify    
\begin{equation}
(1,\operatorname{grad}\tau_g(y))=   |\det D\,S(S^{-1}(y))|^{-1}\,   A\cdot (1,\operatorname{grad}\tau(S^{-1}(y)))\tag{*}
\end{equation} 
which is easy in the case $A\in SU(2)$. So it suffices to check the case $g=\e^{\rho\sigma_3/2}$, $\rho\in\R$.\footnote{ Explicitly  $\e^{\,\rho\,\sigma_3/2}=\operatorname{diag}(\e^{\rho/2},\e^{-\rho/2})$ acts on $\R^4$ by $\left(\begin{array}{cccc}\cosh(\rho) & 0&0&\sinh(\rho)\\ 0&1&0&0\\0&0&1&0\\ \sinh(\rho)&0&0&\cosh(\rho)\end{array}\right)$ for $\rho \in \R$} Put $c:=\cosh \rho$, $s:=\sinh \rho$, $z:=\operatorname{grad} \tau(x)$, $x=S^{-1}(y)$. The rows of  $ \big(D\,S(x)\big)^{-1}$ are $(1,0,0)$, $(0,1,0) $, $\frac{1}{c+sz_3}(-sz_1,-sz_2, 1)$. So the right side of (*) equals $\frac{1}{c+sz_3}(c+sz_3,z_1,z_2,cz_3+s)$. On the left hand side $\operatorname{grad}\tau_g(y))=\big(cz_1-\frac{(cz_3+s)sz_1}{c+sz_3},\dots,\frac{cz_3+s}{c+sz_3}\big)$. Hence (*) holds thus accomplishing the proof.\qed

\section{Localization on achronal not spacelike hyperplanes}

The achronal not spacelike hyperplanes 
$$\kappa:=\{\mathfrak{x}\in\R^4: \mathfrak{x}\cdot \mathfrak{e}=\tau\} \quad \text{ with unique }   \mathfrak{e}=(1,e),\; |e|=1 \text{ and }  \tau\in\R$$
are the tangent spaces to the light cones. They are smooth maximal achronal sets but not Cauchy surfaces for the disjoint  parallel  lightlike straight lines. The main difficulty in defining  a localization on  $\kappa$ via (\ref{TAS}) is the proof of the normalization $\pi_{\phi,\kappa}(\kappa)=||\phi||^2$. The method  in \cite[Proposition 37]{DM24} does not apply without further ado  just because of the existence of the disjoint parallel lightlike straight lines.
\\
\hspace*{6mm}
On the other hand there are  irrefutable physical reasons  for a localization on the achronal not spacelike hyperplanes as briefly expounded in \cite[sec.\,3.4]{C24}. Regarding the Dirac and Weyl fermions see also \cite{C17}. A detailed explanation will be given in sec.\,\ref{PRAL}.
\\
\hspace*{6mm}
Actually,  $\pi_{\phi,\kappa}(\kappa)=||\phi||^2$ holds. Using RKHS  (Reproducing Kernel Hilbert Space) we are going to extend $T$ obeying  (\ref{TAS}) beyond the differentiable Cauchy surfaces in \cite{DM24} to a covariant localization of the massive scalar boson including the achronal not spacelike hyperplanes.

\begin{The}\label{MR} Let $\kappa$ be an achronal not spacelike hyperplane. Then there is a separable Hilbert space $\mathcal{K}$ and an isometry $j:L^2(\R^3)\to L^2(\R^3,\mathcal{K})$ such that
$$\Delta \mapsto T(\Delta):=j^*\mathcal{F}E^{can}(\varpi(\Delta))\mathcal{F}^{-1}j$$
 for every Borel set $\Delta\subset\kappa$, is a localization of the massive scalar boson on $\kappa$ obeying
 \begin{equation*}
\langle \phi,T(\Delta)\phi\rangle=\pi_{\phi,\kappa}(\Delta)
\end{equation*}
 Here $E^{can}$ is the canonical projection valued measure, i.e., $E^{can}(B)\varphi=1_B\varphi$, $B\subset \R^3$ Borel, and $\mathcal{F}$ the Fourier transformation on  $L^2(\R^3,\mathcal{K})$.
 \\
\hspace*{6mm} 
 The extension of $T$ to all Borel subsets $\Delta$ of achronal not spacelike hyperplanes  is Poincar\'e covariant, i.e.,  
 $T(g\cdot\Delta)=W(g)T(\Delta) W(g)^{-1}$  for $g\in \tilde{\mathcal{P}}$.
  \end{The}

The proof of (\ref{MR}) is postponed to the appendix.

\section{High boost limit of spacelike hyperplanes}

The achronal not spacelike hyperplane $\chi=\{\mathfrak{x}\in\R^4: x_0=x_3\}$ is the high boost limit of the Euclidean space $\varepsilon=\{\mathfrak{x}\in\R^4: x_0=0\}$ as follows.  In the same way, by relativistic symmetry, every achronal not spacelike hyperplane is the high boost limit of a spacelike hyperplane. In (\ref{WLB}) and (\ref{MCTC})  we show that 
the localization operators of  $T$ on $\varepsilon$ and  $\chi$ are closely related by the high boost limit.
\\
\hspace*{6mm}
 Recall that $A_\rho:=\e^{\rho\, \sigma_3/2}$ represents the boost along the third spatial axis with rapidity $\rho$. 
 Note $A_\rho\cdot \varepsilon=\{x_0=\tanh(\rho)\, x_3\}$. For $\rho\ge 0$
$$l_\rho:\varepsilon\to A_\rho\cdot \varepsilon, \quad l_\rho(0,x):=\big( \text{\footnotesize{$\frac{1}{2}$}}    (1-\e^{-2\rho})x_3,\,x_1,\, x_2,\, \text{\footnotesize{$\frac{1}{2}$}}    (1+\e^{-2\rho})x_3\big)$$ 
is a linear  bijection composed by the inhomogeneous dilation $\mathfrak{x}\mapsto (x_0,x_1,x_2,\e^{-\rho}x_3)$ and the subsequent boost $\mathfrak{x}\mapsto A_\rho\cdot \mathfrak{x}$. Pointwisely $ l_\rho \to l_\infty$ for $\rho\to\infty$ with 
$$   l_\infty:\varepsilon \to \chi, \quad        l_\infty (0,x):=(\text{\footnotesize{$\frac{1}{2}$}} x_3,x_1,x_2,\text{\footnotesize{$\frac{1}{2}$}} x_3)$$
Hence for $0<\alpha<\beta$ and $0\le \rho <\infty$
\begin{equation}\label{HBL}
l_\rho(\{\mathfrak{x}\in\varepsilon:-\alpha\le x_3 \le \beta\}=\operatorname{e}^{\,\rho\,\sigma_3/2}\cdot\{\mathfrak{x}\in\varepsilon: -\alpha \e^{-\rho}\le x_3\le \beta \operatorname{e}^{-\rho} \}
 \end{equation}
and
\begin{equation}\label{HBLL}
l_\infty(\{\mathfrak{x}\in\varepsilon:-\alpha\le x_3 \le \beta\}=\{\mathfrak{x}\in\chi: -\alpha/2\le  x_3\le\beta/2\}
\end{equation}

The maps $l_\rho$ are charaterized by fact that  every point $(0,x)\in \varepsilon$ runs through the segment $\{l_\rho(0,x): 0\le \rho\le \infty\}$ of the lightlike line $(0,x)+\R(\frac{x_3}{2},0,0,-\frac{x_3}{2})$ joining $(0,x)$ with $(\frac{x_3}{2},x_1,x_2,\frac{x_3}{2})\in\chi$.\\

The localization operators of  $T$ on $\varepsilon$ and  $\chi$ are related to each other by the high boost limit as follows.

\begin{Pro}\label{WLB} Let $\phi$ be a state of the massive scalar boson. Let the Borel set $\Delta\subset \varepsilon$ be bounded. Then  $\lim_{\rho\to\infty} \langle \phi,T\big(l_\rho(\Delta)\big)\phi\rangle=\langle \phi,T\big(l_\infty(\Delta)\big)\phi\rangle$.
\end{Pro}
\\
{\it Proof.} It suffices to prove the claim for $\phi\in C_c$. Put $t_\rho:=\tanh(\rho)$. By (\ref{DETAS}), $\langle \phi,T\big(l_\rho(\Delta)\big)\phi\rangle=\pi_{\phi,A_\rho\cdot \varepsilon}\big(l_\rho(\Delta)\big) $ equals  $\int_{\varpi(l_\rho(\Delta))} \big(J_0(\phi,t_\rho x_3,x)-t_\rho J_3(\phi,t_\rho x_3,x)\big) \d^3x$. Here  for  $\rho\to \infty$ clearly  
$1_{ \varpi(l_\rho(\Delta))} \to 1_{\varpi(l_\infty(\Delta))}$, and by (\ref{FVCCD})  the integrand (without  $ (2\pi)^{-3}$ and the $\phi$-factors)  $\frac{1}{2}\big(\varepsilon(k) + \varepsilon(p) - (k_3+p_3)t_\rho\big) g(\mathfrak{k} \cdot \mathfrak{p} ) \e^{\i (\varepsilon(k) - \varepsilon(p))t_\rho x_3} \e^{\i (p-k)x}$ tends  to $\textsc{k}_\chi(k,p)  \e^{\i\big((p-k)x-(\epsilon(p)-\epsilon(k))x_3\big)} $ (\ref{MOC}). By the assumptions on $\phi$ and $\Delta$ the result follows by dominated convergence.\qed

The  result (\ref{MCTC}) is decisive for deriving  Lorentz contraction for the massive scalar boson. 

\begin{The}\label{MCTC}
Let $J\subset\R$ be an interval (bounded or not bounded, closed or not closed).   Let  $\Gamma\subset\varepsilon$ be the strip $\{x_0=0,x_3\in J\}$. Then

\hspace*{6mm}
\emph{(a)}  $\lim_{\rho\to\infty}T\big(l_\rho(\Gamma)\big) =T\big(l_\infty(\Gamma)\big)$ strongly

\hspace*{6mm}
\emph{(b)}  $T\big(l_{\rho'}(\Gamma)\big)\le  T\big(l_\rho(\Gamma)\big)$ for $0\le \rho \le\rho'\le \infty$ if $0\in\overline{J}$

\hspace*{6mm}
\emph{(c)}  $T(\{x_0=x_3\ge \alpha\}) =  T(\{x_0=\alpha, x_3\ge \alpha\}) $ for $\alpha\in\R$

The equation in \emph{(c)} holds also if $\ge$ is replaced by $>$ or $\le$ or $<$.
\end{The}

The proof  of   (\ref{MCTC}) is postponed to the appendix.

\section{Physical relevance}\label{PRAL}

We will display two consequences of physical relevance of 
the  localization  
on achronal not spacelike hyperplanes. The first is the additivity of the extension $T$, which preserves its normalization and thus
supports the concept of achronal localization as studied  in \cite{C24}. The second concerns the derivation of Lorentz contraction, which apparently for the first time is shown for the massive scalar boson. For the Lorentz contraction regarding the Dirac electron and positron and the four Weyl fermions see \cite{C17}.\\
\hspace*{6mm}
As shown in  \cite[sec.\,11]{C23}, 
 $T$ is causal in the sense that the probability of localization in a \textbf{region of influence} is not less than that in the region $\Delta$ of actual localization. Here $\Delta$ is flat spacelike  measurable. 
 If  $\sigma$ is a spacelike hyperplane, then the (minimal) region of influence $\Delta_\sigma$ of $\Delta$ in $\sigma$  is the set of all points  in $\sigma$, which can be reached  from some point  in $\Delta$ by a signal not moving faster than light. 
Explicitly, $\Delta_\sigma:=\{\mathfrak{x}\in \sigma: \exists \,\mathfrak{y}\in\Delta \text{ with } (\mathfrak{x}-\mathfrak{y})^{\cdot  2}\ge 0\}$. 
So by causality 
\begin{equation}\label{CCFR} 
T(\Delta)\le T(\Delta_\sigma)
\end{equation}
We like to mention the causal localizations regarding the Dirac electron and positron and the four Weyl fermions in \cite{C17}.

\subsection{Normalization}
The question is what about the surplus of spatial probability in $\Delta_\sigma$ with respect to $\Delta$ due to the causality requirement (\ref{CCFR}).

\begin{The}\label{AET}  Let $-\infty<\alpha<\beta<\infty$. Let $\Delta$ be the spacelike half-hyperplane $\{x_0=\alpha, x_3\le\alpha\}$ and $\sigma$ the spacelike hyperplane $\{x_0=\beta\}$,  and $\Gamma:=\{x_0=\beta, x_3>\beta\}$.
 Then 
\begin{equation}\label{ADTLH}
T(X)= T(\Delta_\sigma)-T(\Delta)
\end{equation}
holds for $X:=\{\mathfrak{x}\in\chi:\alpha<x_3\le \beta\}$. Equivalently \begin{equation}\label{AADTLH}
T(\Delta)+T(X)+T(\Gamma)=I
 \end{equation}
for the maximal achronal set  $\Delta\cup X \cup\Gamma$. 
\end{The} 
\\
{\it Proof.} By (\ref{MCTC})(c)  one has $T(\Delta) = T(\{x_0=x_3\le \alpha\})$ and, with (\ref{EEHBL}),  $T(\Gamma) = T(\{x_0=x_3> \beta\})$. Note $\Delta_\sigma=\sigma\setminus \Gamma$. Moreover, 
$\chi=\{x_0=x_3\le \alpha\}\cup X\cup \{x_0=x_3> \beta\}$, where the sets are disjoint. Hence $T(\Delta_\sigma)=I-T(\Gamma)$ and, by  (\ref{MR}), $I=T(\chi)= T(\{x_0=x_3\le \alpha\})+T(X)+T(\{x_0=x_3> \beta\})$, whence the claim.\qed

The relation (\ref{ADTLH}) is not obvious. It is remarkable as it relates localization operators regarding different  spacelike hyperplanes. It allows to interpret
the expectation value $\langle \phi,T(X)\phi\rangle$ as the amount of  probability of localization in $\Delta_\tau$ which for causality  is not due to the localization in $\Delta$. The extension 
of $T$ to achronal not spacelike hyperplanes via  the high boost limit is requested by causality. It guarantees the normalization (\ref{AADTLH})  being fundamental 
for  achronal localization  \cite{C24}.  As already shown in \cite{DM24} normalization   holds   on  differentiable Cauchy surfaces.  
Apparently one is one step prior an achronal localization of the massive scalar boson and hence a representation of the causal logic thus furnishing a complete description  of causality \cite{C24}.

\subsection{Lorentz contraction}

Recall that $A_{\rho e}:=\operatorname{exp}(\frac{\rho}{2}\sum_{k=1}^3e_k\sigma_k)$  represents the boost in direction $e\in\R^3$, $|e|=1$ with rapidity $\rho$. 

\begin{The}\label{LCMSB} Let $\phi$, $||\phi||=1$, be a state of the massive scalar boson.  Then
$$\langle\, W(A_{\rho e})\phi, T(\{\mathfrak{x}\in\varepsilon:-\delta \le xe \le \delta\}) \, W(A_{\rho e})\phi\, \rangle \to 1,\quad |\rho|\to\infty$$
 for every   $\delta>0$.
\end{The}\\
{\it Proof.} It suffices to treat the case $\rho\to\infty$ and every $e$. Indeed, for the case $\rho\to -\infty$ consider $-e$. Then, due to Euclidean covariance,  it suffices to deal  with only one  direction $e$. We choose $e=(0,0,-1)$. 
\\
\hspace*{6mm}
Let $\epsilon>0$. There is $0<\beta<\infty$ such that $\langle \phi, T(\{\mathfrak{x}\in\chi:|x_3|\le \beta/2\}) \,\phi\rangle\ge 1-\epsilon$  since $T(\chi)=I$. By (\ref{HBLL}), $\{\mathfrak{x}\in\chi:|x_3|\le \beta/2\} =l_\infty(\{\mathfrak{x} \in\varepsilon:|x_3|\le \beta\})$. 
Let $\rho_\beta>0$ with $\beta\e^{-\rho}\le \delta$ for $\rho\ge \rho_\beta$. 
\\
\hspace*{6mm}
Then  by (\ref{HBL}) and (\ref{MCTC}), 
$1\ge \langle W(A_{-\rho})\phi, T(\{\mathfrak{x}\in\varepsilon:|x_3| \le \delta \})\, W(A_{-\rho})\phi\rangle \ge \langle W(A_{-\rho})\phi, T(\{\mathfrak{x}\in\varepsilon:|x_3| \le \beta\e^{-\rho}\})\, W(A_{-\rho})\phi\rangle  = \langle \phi, T(A_\rho\cdot \{\mathfrak{x}\in\varepsilon:|x_3| \le \beta\e^{-\rho}\})\,\phi\rangle   \downarrow_\rho \langle \phi, T(\{\mathfrak{x}\in\chi:|x_3|\le \beta/2\}) \,\phi \rangle \ge 1 - \epsilon $. Hence $1\ge  \langle W(A_{-\rho})\phi, T(\{\mathfrak{x}\in\varepsilon:|x_3| \le \delta \})\, W(A_{-\rho})\phi\rangle \ge 1-\epsilon$ for $\rho\ge\rho_\beta$ and every $\epsilon>0$. The result follows.\qed
\\

Hence the  probability of localization  of the boson in the boosted state $W(A_{\rho e})\phi$ in a whatever narrow strip $\{-\delta\le xe\le \delta\}$    tends to $1$ if the rapidity $\rho$ tends  to $\infty$ or $-\infty$. We like to call  this behavior  the \textbf{Lorentz contraction} of  the boson.
\\
\hspace*{6mm} 
Let us  briefly discuss the usual questions related to classical Lorentz contraction. For more details cf. \cite[sec.\,17.2]{C17}.
\\
\hspace*{6mm} 
(a) Can Lorentz contraction be observed?   Immagine an apparatus $\mathcal{A}$ able to ascertain the probabilities of localization of the boson in $\{|xe|\le \delta \}$, i.e.,  the expectation  values of $A:=T(\{|xe|\le \delta \})$. For a given state $S$ described by $\phi$ and $\varepsilon >0$, let $\delta>0$ be so small that $\langle \phi, T(\{|xe|\le \delta \})\phi\rangle \le \varepsilon$. According to (\ref{LCMSB}) there is a rapidity $\tilde{\rho}$ such that 
$\langle\, W(A_{\rho e})\phi, T(|xe| \le \delta\}) \, W(A_{\rho e})\phi\ \rangle \ge 1-\varepsilon$ for $\rho\ge \tilde{\rho}$. Let $\tilde{S}$ be the boosted state. It is described by  $\tilde{\phi}= W(A_{\tilde{\rho} e})\phi$. Then
\begin{equation}\label{LCDSM} 
\langle \phi,A \phi\rangle\le \varepsilon\;\textrm{ and }\;\langle \tilde{\phi},A \tilde{\phi}\rangle\ge 1-\varepsilon
\end{equation}
 Hence the apparatus $\mathcal{A}$ distinguishes the state $S$ from the boosted state $\tilde{S}$. So an observer  can ascertain the Lorentz contraction of the boson.
\\
\hspace*{6mm} 
(b) The ascertainments (\ref{LCDSM}) are related to some reference frame $\mathfrak{R}$. 
What are the ascertainments  of an observer related to any other frame $\mathfrak{R}'\equiv g^{-1}\cdot\mathfrak{R}$
 with $g\in \tilde{\mathcal{P}}$  provided with the  localization $T'$? Note that $A'=T(g\cdot\{x_0=0,|xe|\le\delta\})$ and  $\tilde{\phi}'=W(h')\phi'$ for $h'=ghg^{-1}$, $h:=A_{\tilde{\rho}e}$. For these well-known general relations see e.g. \cite[sec.\,VIII]{A69},  \cite[sec.\,17.2]{C17}. Then 
\begin{equation} 
 \langle \phi',A'\phi'\rangle\le\varepsilon \text{\, and } \langle \tilde{\phi}',A'\tilde{\phi}'\rangle\ge 1-\varepsilon
 \end{equation}
  Hence, observed from $\mathfrak{R}'$, the boson in the state $S$ is highly localized in the spacelike region $g\cdot\{\mathfrak{x}: x_0=0, |xe|>\delta\}$, whereas in the boosted state $\tilde{S}$ it is highly localized in $g\cdot\{\mathfrak{x}:x_0=0, |xe|\le\delta\}$. The expected conclusion is that, due to relativistic symmetry,  the  Lorentz contraction of the massive scalar boson  can be ascertained  in the same way and with the same result by any Lorentz observer.
\\
\hspace*{6mm} 
 (c) On the other hand there is the dependence of the Lorentz contraction on the frame, which is discussed now.  
Due to the relativistic symmetry and the covariance of $T$ the result (\ref{LCMSB}) can be expressed equivalently in the following way. Let a four vector $\mathfrak{e}$ be called a spacelike direction if 
$\mathfrak{e}\cdot \mathfrak{e}=-1$.

\begin{Cor} \label{FDLCDL}  Let $\sigma$ be a spacelike hyperplane and $\mathfrak{e}$ a spacelike direction parallel to $\sigma$. Boost them along $\mathfrak{e}$ with rapidity $\rho$ obtaining 
$\sigma_\rho$ and $\mathfrak{e}_\rho$. Then
$$   \langle\phi,T\big(\{\mathfrak{x}\in\sigma_\rho: |(\mathfrak{x}-\mathfrak{o})\cdot \mathfrak{e}_\rho|\le \delta\}\big)\phi \rangle \to 1 \textrm{ for } |\rho|\to \infty$$ where  $\mathfrak{o}\in\sigma$ is the fixed point of the boost.
\end{Cor}

 Thus, if the frame is moving fast enough depending on the state, then  the boson is highly localized in a narrow strip perpendicular to the direction of  motion.
\\
\hspace*{6mm} 
 The dependence on the frame of the Lorentz contraction in classical  mechanics is striking by the fact that for the comoving observer it does not even exist. The same holds true for the Lorentz contraction of the boson wavefunctions. Moreover, due to  the Poincar\'e covariance of the localization no reference to a moving observer is needed,  but the fact refers to the expectation value of the corresponding localization observable. Indeed, boost the apparatus $\mathcal{A}$ according to  $A_{\tilde{\rho}e}$ thus obtaining the comoving apparatus $\tilde{\mathcal{A}}$. It is able to ascertain  the probabilities of localization of the boson in the space-like region $A_{\rho e}\cdot\{|xe| \le \delta\}$ (to which a comoving observer refers), i.e., the expectation  values of  $\tilde{A}:=T\big(A_{\rho e}\cdot\{|xe| \le \delta\}\big)$. Then due to the covariance of localization
 \begin{equation} \label{NLCFCM}
\langle \tilde{\phi},A \tilde{\phi}\rangle\ge 1-\varepsilon\;\textrm{ and }\;  \langle \tilde{\phi},\tilde{A} \tilde{\phi}\rangle\le \varepsilon
\end{equation} 
holds. This means that the non-comoving apparatus $\mathcal{A}$ ascertains the Lorentz contraction of the boson whereas the comoving apparatus $\tilde{\mathcal{A}}$ ascertains non-contraction.

\appendix

\section{Proof of (\ref{MR}) Theorem}

In order to prove (\ref{MR}) it suffices to deal with the achronal not spacelike hyperplane $$\chi:=\{\mathfrak{x}\in\R^4:x_0=x_3\}$$ due to relativistic symmetry. Indeed, $\kappa=g\cdot \chi$   for the Poincar\'e transformation
 $g=\big((\tau,0),B\big)h$ with $B\in SU(2)$ satisfying $B\cdot (0,0,1)=e$ and arbitrary $h$ 
leaving $\chi$ invariant. Then (\ref{DETAS}) reads
\begin{equation}\label{MOC}
 \pi_{\phi,\chi}(\Delta)=(2\pi)^{-3}\int _{\varpi(\Delta)}\int\int  \textsc{k}_\chi(k,p)   \e^{\i\big((p-k)x-(\epsilon(p)-\epsilon(k))x_3\big)}  \overline{\phi(\mathfrak{k})}\phi(\mathfrak{p})   \, \d o(\mathfrak{k}) \d o(\mathfrak{p})\, d^3 x
 \end{equation}
with $\textsc{k}_\chi(k,p):=\frac{1}{2} \big(\epsilon(k)-k_3+\epsilon(p)-p_3\big)g(\mathfrak{k}\cdot\mathfrak{p})$.

\begin{Lem}  $\textsc{k}_\chi$ is a positive definite kernel on $\R$ with  $\textsc{k}_\chi(p,p)=\epsilon(p)-p_3$.
\end{Lem}\\
{\it Proof.} Let  $Z_B(k,p):=(2\pi)^{-3}\int _B\e^{\i\big((p-k)x-(\epsilon(p)-\epsilon(k))x_3\big)}\d^3 x$ for $B\subset \R^3$  a bounded Borel set. Then $\pi_{\phi,\chi}(\Delta)=\int\int Z_{\varpi(\Delta)}(k,p)\textsc{k}_\chi(k,p) \overline{\phi(\mathfrak{k})}\phi(\mathfrak{p})   \, \d o(\mathfrak{k}) \d o(\mathfrak{p})$ for bounded $\Delta$ and all $\phi\in C_c$. As $\pi_{\phi,\chi}(\Delta)\ge 0$,  
$Z_{\varpi(\Delta)}\textsc{k}_\chi$ is a positive definite kernel on $\R^3$. In particular for $\alpha>0$ one has $Z_{[-\alpha,\alpha]^3}(k,p)=(\alpha/\pi)^3\operatorname{sinc}(\alpha(p_1-k_1)) \operatorname{sinc}(\alpha(p_2-k_2)) \operatorname{sinc}(\alpha(\epsilon(k)-k_3-\epsilon(p)+p_3))$ and $(\pi/\alpha)^3 Z_{[-\alpha,\alpha]^3}\textsc{k}_\chi\to \textsc{k}_\chi$ for $\alpha\to 0$ pointwisely, whence the claim.\qed

\begin{Lem}\label{FPDK} There is a separable Hilbert space $\mathcal{K}$ and  a measurable map $v:\R^3\to \mathcal{K}$ with $||v(p)||=1$  such that $\langle v(k),v(p)\rangle = (\epsilon(k)-k_3)^{-1/2}(\epsilon(p)-p_3)^{-1/2}\textsc{k}_\chi(k,p)$. Obviously $V:L^2(\R^3)\to L^2(\R^3,\mathcal{K})$, $(V\varphi)(p):=v(p)\varphi(p)$ is an isometry.
\end{Lem}\\
{\it Proof.} Let  $\mathcal{K}$ be the RKHS associated to the normalized positive definite kernel $(\epsilon(k)-k_3)^{-1/2}(\epsilon(p)-p_3)^{-1/2}\textsc{k}_\chi(k,p)$.\qed

\begin{Lem}\label{SRTMR} 
$X:L^2(\mathcal{O})\to L^2(\R^3)$,  $(X\phi)(p)=\epsilon(p)^{-1/2}\phi(\mathfrak{p})$ is a Hilbert space isomorphism. 
\end{Lem}\\
{\it Proof.} This is obvious.\qed

\hspace*{6mm}
In the following the change of variables $H$ is used. Put  $\R_-^3:=\R\times \R\times ]-\infty,0[$. Then $H:\R^3\to  R_-^3$, $H(p):=(p_1,p_2,p_3-\epsilon(p))$. $H$ is bijective with $H^{-1}(s)=\big(s_1,s_2, \frac{s_3^2-(m^2+s_1^2+s_2^2)}{2s_3}\big)$. This is easily verified.

\begin{Lem}\label{FFT}  
Let $f\in L^1(\R^3,\mathcal{K})$, $x\in\R^3$.
Then
\begin{equation}
\int_{\R^3}\e^{\i(px-\epsilon(p)x_3)}f(p)\d^3 p= \int_{\R^3_-}\e^{\i sx} \frac{\epsilon(s)^2}{2s_3^2} f(H^{-1}(s))\d^3s\tag{*}
\end{equation}
\end{Lem}\\
{\it Proof.} The left side of (*) equals $L:=\int_{\R^3}\e^{\i H(p)x}f(p)\d^3 p$. 
 Let $\lambda$ be the Lebesgue measure on $\R^3$. Then integration with respect to the image measure $H(\lambda)$ yields $L=\int_{\R^3_-}\e^{\i sx} f(H^{-1}(s))\d H(\lambda)(s)$. Recall  $\d H(\lambda)/\d \lambda =|\det DH^{-1}|$.
The latter equals $s\mapsto \frac{\epsilon(s)^2}{2s_3^2}$, whence the claim.\qed

\begin{Lem} \label{ICVH}
$Y: L^2(\R^3,\mathcal{K}) \to  L^2(\R^3,\mathcal{K}) $, 
$$(Yf)(s):=\left(\frac{\epsilon(H^{-1}(s))}{\epsilon(H^{-1}(s))-H^{-1}(s)_3}\right)^{1/2} 
f\big(H^{-1}(s)\big)$$
 if $s\in\R^3_-$ and $=0$ else, is an isometry.
\end{Lem}\\
{\it Proof.} Note $\big(\d H^{-1}(\lambda)/ \d\lambda\big)(p)=\frac{\epsilon(p)-p_3}{\epsilon(p)}$. Hence the claim follows by integration by substitution.\qed

\begin{Po} \textbf{(\ref{MR})\,Theorem.} Using (\ref{FPDK}),  $ \pi_{\phi,\chi}(\Delta)=\int _{\varpi(\Delta)}\langle R(\phi,x),R(\phi,x)\rangle \d^3x$ for
$(2\pi)^{3/2}R(\phi,x):=\int\e^{\i(px-\epsilon(p)x_3)} \sqrt{\epsilon(p)-p_3 \,}\phi(\mathfrak{p}) \,v(p)\,\d o(\mathfrak{p})$. Now applying (\ref{SRTMR}), (\ref{FPDK}), (\ref{FFT}), (\ref{ICVH}) in turn one gets  $(2\pi)^{3/2}R(\phi,x)= \int \e^{\i(px-\epsilon(p)x_3)}   
\sqrt{\frac{\epsilon(p)-p_3}{\epsilon(p)}}   
 (X \phi)(p) 
 v(p)\,\d^3p = 
  \int \e^{\i(px-\epsilon(p)x_3)} \sqrt{\frac{\epsilon(p)-p_3}{\epsilon(p)}}    (VX \phi)(p) \,\d^3p 
 =  \int 1_{\R^3_-}(s)
\e^{\i sx} \frac{\epsilon(s)^2}{2s_3^2} 
\Big(\frac{\epsilon(H^{-1}(s))-H^{-1}(s)_3}{\epsilon(H^{-1}(s))}\Big)^{1/2} \\
(VX\phi)\big(H^{-1}(s)\big) \d^3s =   \int 1_{\R^3_-}(s)
\e^{\i sx} 
\Big(\frac{\epsilon(H^{-1}(s))}{\epsilon(H^{-1}(s))-H^{-1}(s)_3}\Big)^{1/2} 
(VX\phi)\big(H^{-1}(s)\big) \d^3s =  \int 
\e^{\i sx} 
(YVX\phi)(s)\d^3s = (2\pi)^{3/2} (\mathcal{F}^{-1}(YVX\phi))(x)$.
\\
\hspace*{6mm}
So $j:L^2(O)\to L^2(\R^3,\mathcal{K})$, $j:=YVX$ is an isometry. One has $\pi_{\phi,\chi}(\Delta)=\int \langle( \mathcal{F}^{-1}j\phi)(x), 1_{\varpi(\Delta)}(x)(\mathcal{F}^{-1}j\phi)(x)\rangle \d^3x =  \langle \mathcal{F}^{-1}j\phi, E^{can}(\varpi(\Delta))\mathcal{F}^{-1}j\phi \rangle$. The proof of the first part of the assertion  is easily accomplished.
\\
\hspace*{6mm}
The covariance of $T$ follows from (\ref{CPOL})(b). Note that  $T_\kappa(\Delta)=T_{\kappa'}(\Delta)=0$ holds in the case  of $\Delta\subset \kappa\cap\kappa'$ for $\kappa\ne\kappa'$  as $\varpi(\kappa\cap\kappa')$ is a Lebesgue null set.
\qed
\end{Po}

\section{Proof of (\ref{MCTC}) Theorem}

For every subset $M$ of Minkowski spacetime   let $ M^{\sim}$ denote its \textbf{set of determinacy}, i.e., the set of  all events $\mathfrak{x}$ such that every timelike straight  line through $\mathfrak{x}$  meets $M$. Recall that $\Gamma_\sigma$ denotes the region of influence of the region $\Gamma$ in the spacelike hyperplane $\sigma$. 
 
\begin{Lem}\label{BODS} 
 Let $\sigma, \tau$ be  spacelike hyperplanes and let $\Delta\subset\sigma$ be measurable.  Further let $\Gamma\subset \Delta^{\sim}\cap \tau$ be measurable. 
 Then $T(\Gamma)\le T(\Delta)$.
 If $\Gamma^{\sim}=  \Delta^{\sim}$  then $T(\Gamma)=T(\Delta)$.
  \end{Lem}\\
 {\it Proof.} Obviously $M(\Gamma,\sigma):=\bigcup_{\mathfrak{y}\in\Gamma}\{\mathfrak{x}\in\sigma: (\mathfrak{x}-\mathfrak{y})^{\cdot 2}>0\} $ is open and contained in $\Gamma_\sigma$.   By \cite[(16) Lemma]{C17} it equals $\Gamma_\sigma$ up to a null set. 
Now the claim is  $M(\Gamma,\sigma)\subset \Delta$. Then  causality (\ref{CCFR})   implies $T(\Gamma)\le T(\Gamma_\sigma)=T(M(\Gamma,\sigma))\le T(\Delta)$.  In order to get the last part of the assertion interchange the roles of $\sigma,\Delta$ and $\tau,\Gamma$. \\
\hspace*{6mm} 
So let $\mathfrak{x}\in M(\Gamma,\sigma)$. Then there is $\mathfrak{y}\in \Gamma$ with $\mathfrak{z}^{\cdot2}>0$ for $\mathfrak{z}:=\mathfrak{x}-\mathfrak{y}$. Since $\Gamma\subset \Delta^{\sim}$ there is $s\in\R$ such that $\mathfrak{x}':=\mathfrak{y}+s\mathfrak{z}\in\Delta$. Since $\mathfrak{x'}, \mathfrak{x}\in\sigma$, one has 
$(s-1)^2\mathfrak{z}^{\cdot 2}= (\mathfrak{x'}-\mathfrak{x})^{\cdot 2}\le 0$. This requires $s=1$, whence the claim $\mathfrak{x}\in\Delta$.\qed

\begin{Lem}\label{EHBL} Let $\beta\in[0,\infty[$. Consider the case $J=[0,\beta]$ or $J=[0,\infty[$ or  $J=[-\beta,0]$ or $J=]-\infty,0]$.  Put $\Gamma_\rho:=l_\rho(\Gamma)$ for the strip $\Gamma$. Then
$T(\Gamma_{\rho'})\le T(\Gamma_{\rho})$ for $0\le \rho\le\rho'<\infty$, and there is a positive operator  $R(\Gamma_\infty)$ such that $T(\Gamma_\rho)\to R(\Gamma_\infty)$  strongly for $\rho\to \infty$. Finally, for every $0\le v<1$,
$R(\{x_0=x_3\ge 0\})=T(\{x_0=vx_3, x_3\ge 0\}), \; R(\{x_0=x_3\le 0\})=T(\{x_0=vx_3, x_3\le 0\})$ holds.
\end{Lem}\\
{\it Proof.} Let $\varsigma\in\{1,-1\}$. Check  $$\Gamma^\sim_{\rho}=\{\mathfrak{x}:-\beta\e^{-2\rho}\le \varsigma(x_0-x_3)\le 0, 0\le \varsigma(x_0+x_3)\le \beta\}$$ Indeed,  $\Gamma^\sim_{\rho}=\operatorname{e}^{\,\rho\,\sigma_3/2}\cdot\{x_0=0, 0\le \varsigma x_3\le \beta \operatorname{e}^{-\rho} \}^\sim=   \operatorname{e}^{\,\rho\,\sigma_3/2}\cdot\{ -\beta\e^{-\rho}\le x_0-\varsigma x_3\le0, 0\le x_0+\varsigma x_3\le \beta\e^{-\rho}\} =\{\mathfrak{x}: -\beta\e^{-\rho}\le x'_0-\varsigma x'_3\le0, 0\le x'_0+\varsigma x'_3\le \beta\e^{-\rho}\}$ for $x_0':= \cosh(\rho)x_0-\sinh(\rho)x_3$,   $x_3':=- \sinh(\rho)x_0+\cosh(\rho)x_3$, whence the claim. It comprises the cases of infinite $J$ putting $\beta=\infty$.\\
 \hspace*{6mm}  
Now note $\Gamma_{\rho'}\subset \Gamma_{\rho'}^\sim\subset      \Gamma^\sim_{\rho}$ for $0\le\rho\le\rho'<\infty$. Therefore (\ref{BODS}) applies, whence $T(\Gamma_{\rho'})\le T(\Gamma_{\rho})$. The limit $\rho\to \infty$ exists by \cite[Satz 4.28]{W76}. \\
 \hspace*{6mm} 
For $\Gamma= \{x=0,\,x_3\ge 0\}$ or  $\Gamma= \{x=0,\,x_3\le 0\}$  note  $\Gamma_{\rho'}\subset \Gamma_{\rho'}^\sim =     \Gamma^\sim_{\rho}$ for $0\le\rho\le\rho'<\infty$. Put $v:=\tanh \rho$. Hence by 
 (\ref{BODS}) the proof is completed. \qed
 
\begin{Lem} \label{EEHBL} 
 Let $\Gamma'_{\rho}$, $0\le\rho\le\infty$ denote the set $\Gamma_{\rho}$  in \emph{(\ref{EHBL})} with possibly one or both  $\le$ replaced by $<$.  Then  $T( \Gamma'_{\rho})\to R(\Gamma_\infty'):=R(\Gamma_\infty)$.
\end{Lem}\\
{\it Proof.} Since
$\Gamma_{\rho}\setminus\Gamma'_{\rho}$, $\rho<\infty$ is a Lebesgue null set, $T( \Gamma'_{\rho})=T(\Gamma_{\rho})$ holds. Check $\Gamma'^\sim_{\rho'}\subset \Gamma'^\sim_\rho$, $0\le\rho\le\rho'<\infty$. Apply the proof of (\ref{EHBL}).\qed

\begin{Cor}\label{MLT} Let $\Gamma$ be the strip in $\varepsilon$ associated to the interval $J\subset\R$. Let $0\in\overline{J}$. Then $T\big(l_{\rho'}(\Gamma)\big)\le  T\big(l_\rho(\Gamma)\big)$ for $0\le \rho \le\rho'\le \infty$ and  there is a positive operator  $R\big(l_\infty(\Gamma)\big)$ such that  $\lim_{\rho\to\infty}T\big(l_\rho(\Gamma)\big) =R\big(l_\infty(\Gamma)\big)$ strongly.
\end{Cor}\\
{\it Proof.} The strips associated to the intervals $J_-:=J\cap ]-\infty,0[$, $J_+:=J\cap [0,\infty[$ are of the kind treated in (\ref{EEHBL}). The proof is easily accomplished.\qed

\begin{Cor}\label{EHBLS} Let $\Gamma$ be the strip in $\varepsilon$ associated to the interval $J\subset\R$. Then  $\lim_{\rho\to\infty}T\big(l_\rho(\Gamma)\big) =R\big(l_\infty(\Gamma)\big)$ strongly  for some positive operator  $R\big(l_\infty(\Gamma)\big)$.
\end{Cor}\\
{\it Proof.} Obviously it suffices to show the claim for the intervals $J_-:=J\cap ]-\infty,0[$, $J_+:=J\cap [0,\infty[$. Write $J_+=J_2\setminus J_1$ with $J_1\subset J_2$, where the interval $J_i\subset [0,\infty[$, $i=1,2$ is of the kind treated in (\ref{MLT}). The result follows for $J_+$ by  (\ref{MLT}). In the same way  it follows for $J_-$.\qed\\

 \begin{Pro} $T\big(l_\infty(\Gamma)\big) =R\big(l_\infty(\Gamma)\big)$ holds for every strip $\Gamma$.
\end{Pro}
\\
{\it Proof.} (a)  First we show $T\big(l_\infty(\Gamma)\big) \le R\big(l_\infty(\Gamma)\big)$. Let $\Delta\subset \Gamma$ for $\Delta$ in (\ref{WLB}). Put $\Delta_\rho:=l_\rho(\Delta)$.
$T(\Delta_\rho)\le T(\Gamma_\rho)$ for $\rho<\infty$ since $\Delta_\rho\subset \Gamma_\rho$. Assume $\langle \phi,T(\Delta_\infty) \phi\rangle >\langle \phi, R(\Gamma_\infty) \phi \rangle$. There is $\rho_n\to\infty$ with  $\langle \phi,T(\Delta_{\rho_n}) \phi\rangle > \langle \phi, R(\Gamma_\infty) \phi \rangle$ as $\ \langle \phi,T(\Delta_{\rho}) \phi\rangle \to \langle \phi, T(\Delta_\infty) \phi \rangle$ by (\ref{WLB}). So $0\le  \langle \phi,T(\Gamma_{\rho_n}) \phi\rangle -  \langle \phi,T(\Delta_{\rho_n}) \phi\rangle \to  \langle \phi, R(\Gamma_\infty) \phi \rangle -  \langle \phi, T(\Delta_\infty) \phi \rangle$ contradicting the assumption. --- Now approximate $\langle \phi,T(\Gamma_\infty) \phi\rangle$ by $\langle \phi,T(\Delta_\infty) \phi\rangle$  with bounded $\Delta$. The claim follows.
\\
\hspace*{6mm} (b) As for the proof of (\ref{EHBLS}) it suffices to treat the case of $\Gamma=\{x_0=0,0\le x_3\le \beta\}$. The case $\Gamma=\Lambda$ for   $\Lambda:=\{x_0=0, x_3\ge 0\}$ is already shown in (\ref{EHBL}). By (a), $T(\Gamma_\infty)\le R(\Gamma_\infty)$ and $T\big((\Lambda\setminus\Gamma)_\infty\big) \le R\big((\Lambda\setminus\Gamma)_\infty\big)$. Note $(\Lambda\setminus\Gamma)_\rho=\Lambda_\rho\setminus \Gamma_\rho$ for $0\le\rho\le\infty$. Hence $T\big((\Lambda\setminus\Gamma)_\infty\big) =T(\Lambda_\infty)-T(\Gamma_\infty)$ and, by (\ref{EHBLS}),  $R\big((\Lambda\setminus\Gamma)_\infty\big) =R(\Lambda_\infty)-R(\Gamma_\infty)$. Since $R(\Lambda_\infty) = T(\Lambda_\infty)$ by
(\ref{EHBL}),  this implies $T(\Gamma_\infty)\ge R(\Gamma_\infty)$, whence the result.\qed

 The proof of the theorem (\ref{MCTC}) is completed showing now the claim (c).

\begin{Pro}  $T(\{x_0=x_3\sim \alpha\}) =  T(\{x_0=\alpha, x_3\sim \alpha\}) $ holds for $\alpha\in\R$ and $\sim\;\in\{\ge,>,\le,<\}$.
\end{Pro}\\
 {\it Proof.}  Let $\sim\;=\;\ge$.     Put $\mathfrak{e}:=(1,0,0,1)$. By covariance in  (\ref{MR}) and by  (\ref{EHBL}) one has $T(\{x_0=x_3\ge \alpha\}) = T(\alpha\mathfrak{e}+\{x_0=x_3\ge 0\})= W(\alpha\mathfrak{e}) T(\{x_0=x_3\ge 0\}) W(\alpha\mathfrak{e})^{-1} = 
  W(\alpha\mathfrak{e}) T(\{x_0=0, x_3\ge 0\}) W(\alpha\mathfrak{e})^{-1}   =   T(\{x_0=\alpha, x_3\ge \alpha\})$. The other $\sim$\,-cases are analogous using (\ref{EEHBL}). \qed
\\

\textbf{Acknowledgement}.  I am very grateful   to Valter Moretti and Carmine De Rosa   for many valuable discussions. 
\\

\textbf{Competing interests.} The author states that there is no conflict of interests.

\textbf{Data availability.} Data sharing is not applicable to this article as no datasets were generated or analyzed during the current study.

\vspace{2cm}
%For JMP:

%Significance:
%The present investigation yields an irrefutable argument promoting achronal localization as the conceptual frame which complies most completely with the principle of causality for quantum mechanical systems.

%Originality:
%It is shown that causality implies the existence of the high boost limit which relates the probability of localization in acronal not spacelike regions with that in spacelike regions. An application of this relation is the derivation of the Lorentz contraction.

%Advances in the field:
%The considerations provide a more profound understanding of causality and  its possible implications.

\end{document}